\documentclass{article}
\usepackage{amsmath}
\usepackage{epsfig}
\usepackage{amssymb}
\usepackage{fullpage}

\def\##1{{\underline #1}}
\def\=#1{\underline{\underline{#1}}}

\def\+#1{\underline{\bf #1}}
\def\*#1{\underline{\underline{\bf #1}}}

\def\eps{\epsilon}
\def\epso{\epsilon_{\scriptscriptstyle 0}}
\def\muo{\mu_{\scriptscriptstyle 0}}

\def\.{\mbox{ \tiny{$^\bullet$} }}

\def\le{\left(}
\def\ri{\right)}

\def\l#1{\label{#1}}
\def\r#1{(\ref{#1})}
\def\c#1{\cite{#1}}

\def\eps{\epsilon}
\def\epso{\epsilon_0}
\def\muo{\mu_0}

\def\.{\mbox{ \tiny{$^\bullet$} }}

\def\le{\left(}
\def\ri{\right)}

\def\l#1{\label{#1}}
\def\r#1{(\ref{#1})}

\begin{document}

\begin{center} {\bf {\Large Negative refraction by quantum vacuum}}\\
\end{center}
\vskip 1 cm

\noindent {\sf Tom G. Mackay\footnote{Permanent address: School of
Mathematics, James Clerk Maxwell Building, University of Edinburgh,
Edinburgh, EH9 3JZ, UK. Email: T.Mackay@ed.ac.uk}} and
 {\sf Akhlesh Lakhtakia\footnote{Email: akhlesh@psu.edu}} \\
212 Earth \& Engineering Sciences Building\\
Pennsylvania State University\\
University Park\\ PA 16802--6812\\ USA


\vspace{25mm}

\begin{abstract}
The phase velocity of light is co--parallel to the direction of
energy flow in classical vacuum. However, in certain uncommon
materials,  these two vectors can be oppositely directed, in which
case the phase velocity is termed `negative'. This negative phase
velocity (NPV) gives rise to many exotic phenomenons, such as
negative refraction, inverse Doppler shift and inverse \u{C}erenkov
radiation, and has technological allure. According to quantum
electrodynamics,  the presence of a magnetostatic field makes vacuum
an anisotropic medium for the passage of light. Under the influence
of a sufficiently strong
  magnetostatic field, vacuum supports NPV.
Such ultrastrong magnetic fields are believed to arise due to dynamo
action in newborn neutron stars  and in binary neutron star mergers,
for examples. In view of the possible occurrence of negative
refraction, the influence of ultrastrong magnetostatic fields  must
be carefully taken into account in astronomical observations
relating to neutron stars and associated gamma--ray bursts. \vskip 1
cm

\end{abstract}

\vspace{10mm}

In the usual descriptions of light propagation, as learned by
generations of students from  standard textbooks on optics \c{BW}
and electromagnetics \c{Jackson}, the phase velocity casts a
positive projection onto the direction of energy flow, as provided
by the time--averaged Poynting vector. That is, the phase velocity
is \emph{positive}. However, in certain special circumstances, it is
possible for the phase velocity to cast a negative projection onto
the time--averaged Poynting vector; i.e., the phase velocity can be
\emph{negative}.

The transition from positive phase velocity (PPV) to  negative phase
velocity (NPV) is of particular  interest because NPV underpins the
much--heralded phenomenon of negative refraction \c{SSS,Rama}, as
well as many other exotic phenomenons such as inverse Doppler shift
and inverse \u{C}erenkov radiation \c{Veselago68}. The scientific
and technological possibilities offered by negative
refraction~---~such as the
 construction of near--perfect lenses from planar slabs of NPV--supporting
materials~---~have been widely reported on \c{Rama,Pendry}. Much of
the research effort has been directed towards the realization of
\emph{metamaterials} which can be used to achieve negative
refraction. These metamaterials are artificial composite materials,
often  having complex micromorphologies \c{SSS,Dolling,Shalaev}.
Metamaterials with simple micromorphologies can also support NPV, as
is demonstrated by certain homogenized composite materials
 \c{NPV_HCM3}. A  significant milestone was reached recently
   by experimentalists through the fabrication of
metamaterials which support negative refraction at the optical
wavelengths \c{Dolling,Shalaev}.

We report here on the manifestation of NPV in a quite different context,
namely in   vacuum under the influence of a
magnetostatic field $\#B_s= |\#B_s| \,\hat{\#B}_s$. In classical
vacuum, the phase velocity is positive and the passage of light
 is unaffected by $\#B_s$, as reported by an inertial observer.
However, this is not the case for the quantum electrodynamical (QED)
vacuum. The QED vacuum is a nonlinear  medium which can
be linearized for rapidly time--varying plane waves. Thereby, for propagation of
light,  QED
vacuum is represented by the anisotropic dielectric--magnetic
constitutive relations \c{Adler71}
\begin{equation}
\#D = \epso \=\eps \. \#E, \qquad \#B = \muo \=\mu \. \#H, \l{CRs}
\end{equation}
with $\epso=8.854\times10^{-12}$~F~m$^{-1}$ and $\muo=4\pi\times
10^{-7}$~H~m$^{-1}$ being the permittivity and permeability of
 classical vacuum, respectively. The relative permittivity and
 relative permeability dyadics of QED vacuum have the uniaxial forms
\begin{equation}
\left. \begin{array}{l} \=\eps = \le 1-8 a |\#B_s|^2 \ri  \le \=I-
\hat{\#B}_s \hat{\#B}_s \ri + \le 1+20 a |\#B_s|^2 \ri \hat{\#B}_s
\hat{\#B}_s \vspace{8pt} \\ \=\mu = \displaystyle{\frac{1}{ 1-8 a
|\#B_s|^2 } \le \=I- \hat{\#B}_s \hat{\#B}_s \ri + \frac{1}{ 1-24 a
|\#B_s|^2 } \hat{\#B}_s \hat{\#B}_s}
\end{array} \right\}, \l{CDs}
\end{equation}
respectively,  where $\=I$ is the 3$\times$3 identity dyadic and the
constant $a = 6.623 \times 10^{-26}\, {\mbox H}^{-1} {\mbox
{kg}}^{-1}\,\,{\mbox{m}}^2\,\,{\mbox{s}}^2$.
The constitutive
dyadics \r{CDs} were derived by Adler \c{Adler71} from the
Heisenberg--Euler effective Lagrangrian of the electromagnetic field
\c{HE,Schwinger}.

Suppose that plane waves with field phasors
\begin{equation}
 \#E(\#r) = \#E_0 \exp \le i \#k \. \#r \ri, \qquad \#H(\#r) =
\#H_0 \exp \le i \#k \. \#r \ri \l{PWs}
\end{equation}
and wavevector $\#k = | \#k |\, \hat{\#k}$ propagate through the QED
vacuum described by \r{CRs} and \r{CDs}. For simplicity, let
$\hat{\#B}_s$ and $\hat{\#k}$ both be aligned with the Cartesian $z$
axis. In this case there is only one wavenumber, namely $| \#k | =
\omega \sqrt{\epso \muo}\,$ (unlike the $\hat{\#B}_s \neq \hat{\#k}$
case where there are two wavenumbers for each propagation direction,
as is described in the {\bf Supplementary Information} section). The
time--averaged Poynting vector $\#P$ and phase velocity $\#v_p$ are
straightforwardly derived by combining \r{CRs}, \r{CDs} and \r{PWs}
with the Maxwell curl postulates \c{Chen}. Thus, the projection of
the phase velocity onto the time--averaged Poynting vector emerges
as
\begin{equation}
\#v_p \. \#P = \frac{1}{2 \muo} \le 1 - 8 a |\#B_s|^2 \ri | \#E_0
|^2.
\end{equation}
Clearly, the phase velocity is positive for $|\#B_s|< 1/\sqrt{8
a}\,$ but negative for $|\#B_s| > 1/\sqrt{8 a}$. In fact, the same
inequalities  for PPV and NPV hold when the direction of planewave
propagation is perpendicular to the direction of the  magnetostatic
field, as is described in the {\bf Supplementary Information}
section.

Magnetic fields of magnitude greater than $1/\sqrt{8 a} = 1.374
\times 10^{12}$ Tesla are needed for QED vacuum to support NPV
propagation. To find fields of this magnitude we turn to
astrophysical environments. Ultrastrong magnetic fields, developed
by dynamo action, are associated with certain neutron stars. For
example, fields of
 the order of $ 10^{10}$--$ 10^{12}$ Tesla, have been estimated
for newborn neutron stars,  such as  soft gamma repeaters
\c{TD92,TD96}. Considerably stronger fields, of the order of $
10^{11}$--$ 10^{14}$ Tesla, are predicted to arise  during the
merger of a binary neutron star system \c{Price}. Accordingly, NPV
propagation may be expected to occur in these neutron--star
environments. Across boundaries between two regions, one of which
experiences a magnetostatic field of magnitude  $1/\sqrt{8 a} <
1.374 \times 10^{12}$ Tesla and the other a magnetostatic field of
magnitude $1/\sqrt{8 a} > 1.374 \times 10^{12}$ Tesla, negative
refraction of light will occur.

Our results have far--reaching implications for observational
astronomy and theoretical astrophysics. The possibility of negative
refraction arising from ultrastrong magnetic fields should be taken
into consideration in estimating astronomical positions,
particularly if the light path from the detector to the point of
observation traverses regions in the vicinity of neutron stars, for
example. Furthermore, negative refraction may well influence the
propagation of gamma--ray bursts and other electromagnetic radiation
emitted by  neutron stars.

\vspace{10mm}

\noindent{\bf Acknowledgement} TGM is supported by a \emph{Royal
Society of Edinburgh/Scottish Executive Support Research
Fellowship}.

\vspace{10mm}



\vspace{10mm}

\section*{Supplementary Information}

In our Letter we demonstrate that electromagnetic plane waves can
propagate in a quantum electrodynamical (QED) vacuum with negative
phase velocity (NPV), provided that a sufficiently strong
magnetostatic field  is acting.  Here we provide further details of
the planewave analysis. As in our Letter, suppose that plane waves
with field phasors \r{PWs} propagate through the QED vacuum
described by \r{CRs} and \r{CDs}. The wavevector  $\#k = | \#k |
\hat{\#k}$ and corresponding phasor amplitude $\#E_0 = \le E^x_0,
E^y_0, E^z_o \ri$ (and similarly $\#H_0$) are straightforwardly
deduced by combining the frequency--domain constitutive relations
\r{CRs} with the source--free Maxwell curl postulates
\begin{equation}
\left.
\begin{array}{l}
 \nabla \times \#H(\#r) + i \omega \#D(\#r) = \#0 \vspace{2mm} \\
\nabla \times \#E(\#r) - i \omega \#B(\#r) = \#0
\end{array}
\right\}.
\end{equation}

For our purpose, it suffices to
 consider only two cases: (i) propagation parallel to $\hat{\#B}_s$
 and (ii) propagation perpendicular to $\hat{\#B}_s$. Without loss
 of generality, our coordinate system is oriented such that
 $\hat{\#B}_s = \hat{\#z}$.

 For the case $\hat{\#k} = \hat{\#z}$, the dispersion relation
yields one wavenumber, namely $| \#k | = \omega \sqrt{\epso \muo}$.
The electric field phasor lies in the  Cartesian $xy$ plane but is
otherwise arbitrary. The scalar product of the phase velocity and
time--averaged Poynting vector is given as
\begin{equation}
\#v_p \. \#P = \frac{1}{2 \muo} \le 1 - 8 a |\#B_s|^2 \ri \le |
E^x_0 |^2 + | E^y_0 |^2 \ri.
\end{equation}

The case where $\hat{\#k} = \hat{\#x}$ is more complicated as there
are two possible wavenumbers, namely $| \#k | = k_{1,2}$ where
\begin{equation}
k_1 = \omega \sqrt{\epso \muo} \sqrt{\frac{1 - 20 a |\#B_s|^2}{1 - 8
a |\#B_s|^2}}, \qquad k_2 = \omega \sqrt{\epso \muo} \sqrt{\frac{1 -
8 a |\#B_s|^2}{1 - 24 a |\#B_s|^2}}.
\end{equation}
The wavenumber $k_1$ corresponds to a propagating mode for $|\#B_s|
< 1/\sqrt{20a}$ or $|\#B_s| > 1/\sqrt{8a}$; otherwise it is an
evanescent mode. Similarly, the wavenumber $k_2$ corresponds to a
propagating mode for $|\#B_s| < 1/\sqrt{24a}$ or $|\#B_s| >
1/\sqrt{8a}$; otherwise it is an evanescent mode. The electric field
phasor is directed along the  Cartesian $z$ axis for the wavenumber
$k_1$, whereas it is directed along the Cartesian $y$ axis for the
wavenumber $k_2$. The scalar product of the phase velocity and
time--averaged Poynting vector is given as
\begin{equation}
\#v_p \. \#P = \left\{
\begin{array}{lll}
\displaystyle{\frac{1}{2 \muo} \le 1 - 8 a |\#B_s|^2 \ri  | E^z_0
|^2 }&
\mbox{for} & | \#k | = k_{1} \vspace{4mm} \\
\displaystyle{\frac{1}{2 \muo} \le 1 - 24 a |\#B_s|^2 \ri  | E^y_0
|^2 }& \mbox{for} & | \#k | = k_{2}
\end{array} .
\right.
\end{equation}

Therefore,  plane waves propagate with NPV in  QED vacuum provided
that $|\#B_s|
> 1/\sqrt{8 a}$, regardless of whether the propagation direction is
parallel or perpendicular to $\#B_s$.

\end{document}